\documentstyle[twoside,psfig]{article}

\catcode`\@=11
\long\def\@makefntext#1{
\protect\noindent \hbox to 3.2pt {\hskip-.9pt  
$^{{\eightrm\@thefnmark}}$\hfil}#1\hfill}		

\def\@makefnmark{\hbox to 0pt{$^{\@thefnmark}$\hss}}	
	
\def\ps@myheadings{\let\@mkboth\@gobbletwo
\def\@oddhead{\hbox{}
\rightmark\hfil\eightrm\thepage}   
\def\@oddfoot{}\def\@evenhead{\eightrm\thepage\hfil
\leftmark\hbox{}}\def\@evenfoot{}
\def\sectionmark##1{}\def\subsectionmark##1{}}

\oddsidemargin=\evensidemargin
\newcounter{sectionc}\newcounter{subsectionc}\newcounter{subsubsectionc}
\renewcommand{\section}[1] {\vspace{12pt}\addtocounter{sectionc}{1} 
\setcounter{subsectionc}{0}\setcounter{subsubsectionc}{0}\noindent 
	{\tenbf\thesectionc. #1}\par\vspace{5pt}}
\renewcommand{\subsection}[1] {\vspace{12pt}\addtocounter{subsectionc}{1} 
	\setcounter{subsubsectionc}{0}\noindent 
	{\bf\thesectionc.\thesubsectionc. {\kern1pt \bfit #1}}\par\vspace{5pt}}
\renewcommand{\subsubsection}[1] {\vspace{12pt}\addtocounter{subsubsectionc}{1}
	\noindent{\tenrm\thesectionc.\thesubsectionc.\thesubsubsectionc.
	{\kern1pt \tenit #1}}\par\vspace{5pt}}
\newcommand{\nonumsection}[1] {\vspace{12pt}\noindent{\tenbf #1}
	\par\vspace{5pt}}

\newcounter{appendixc}
\newcounter{subappendixc}[appendixc]
\newcounter{subsubappendixc}[subappendixc]
\renewcommand{\thesubappendixc}{\Alph{appendixc}.\arabic{subappendixc}}
\renewcommand{\thesubsubappendixc}
	{\Alph{appendixc}.\arabic{subappendixc}.\arabic{subsubappendixc}}

\renewcommand{\appendix}[1] {\vspace{12pt}
        \refstepcounter{appendixc}
        \setcounter{figure}{0}
        \setcounter{table}{0}
        \setcounter{lemma}{0}
        \setcounter{theorem}{0}
        \setcounter{corollary}{0}
        \setcounter{definition}{0}
        \setcounter{equation}{0}
        \renewcommand{\thefigure}{\Alph{appendixc}.\arabic{figure}}
        \renewcommand{\thetable}{\Alph{appendixc}.\arabic{table}}
        \renewcommand{\theappendixc}{\Alph{appendixc}}
        \renewcommand{\thelemma}{\Alph{appendixc}.\arabic{lemma}}
        \renewcommand{\thetheorem}{\Alph{appendixc}.\arabic{theorem}}
        \renewcommand{\thedefinition}{\Alph{appendixc}.\arabic{definition}}
        \renewcommand{\thecorollary}{\Alph{appendixc}.\arabic{corollary}}
        \noindent{\tenbf Appendix \theappendixc #1}\par\vspace{5pt}}
\newcommand{\subappendix}[1] {\vspace{12pt}
        \refstepcounter{subappendixc}
        \noindent{\bf Appendix \thesubappendixc. {\kern1pt \bfit #1}}
	\par\vspace{5pt}}
\newcommand{\subsubappendix}[1] {\vspace{12pt}
        \refstepcounter{subsubappendixc}
        \noindent{\rm Appendix \thesubsubappendixc. {\kern1pt \tenit #1}}
	\par\vspace{5pt}}

\newcommand{\textlineskip}{\baselineskip=13pt}
\newcommand{\smalllineskip}{\baselineskip=10pt}

\newcommand{\publisher}[2]{{\begin{center}\footnotesize\smalllineskip 
	Received #1\\
	Revised #2
	\end{center}
	}}

\def\abstracts#1#2#3{{
	\centering{\begin{minipage}{6.0in}\footnotesize\baselineskip=10pt
	\parindent=0pt #1\par 
	\parindent=15pt #2\par
	\parindent=15pt #3
	\end{minipage}}\par}} 

\def\keywords#1{{
	\centering{\begin{minipage}{6.0in}\footnotesize\baselineskip=10pt
	{\footnotesize\it Keywords}\/: #1
	\end{minipage}}\par}}

\renewenvironment{thebibliography}[1]
        {\frenchspacing
	 \ninerm\baselineskip=11pt
         \begin{list}{\arabic{enumi}.}
        {\usecounter{enumi}\setlength{\parsep}{0pt}     
	 \setlength{\leftmargin 12.7pt}{\rightmargin 0pt} 
         \setlength{\itemsep}{0pt} \settowidth
	{\labelwidth}{#1.}\sloppy}}{\end{list}}

\newcounter{itemlistc}
\newcounter{romanlistc}
\newcounter{alphlistc}
\newcounter{arabiclistc}

\newcommand{\fcaption}[1]{
        \refstepcounter{figure}
        \setbox\@tempboxa = \hbox{\footnotesize Fig.~\thefigure. #1}
        \ifdim \wd\@tempboxa > 5in
           {\begin{center}
        \parbox{6in}{\footnotesize\smalllineskip Fig.~\thefigure. #1}
            \end{center}}
        \else
             {\begin{center}
             {\footnotesize Fig.~\thefigure. #1}
              \end{center}}
        \fi}

\newcommand{\tcaption}[1]{
        \refstepcounter{table}
        \setbox\@tempboxa = \hbox{\footnotesize Table~\thetable. #1}
        \ifdim \wd\@tempboxa > 5in
           {\begin{center}
        \parbox{6in}{\footnotesize\smalllineskip Table~\thetable. #1}
            \end{center}}
        \else
             {\begin{center}
             {\footnotesize Table~\thetable. #1}
              \end{center}}
        \fi}

\def\@citex[#1]#2{\if@filesw\immediate\write\@auxout
	{\string\citation{#2}}\fi
\def\@citea{}\@cite{\@for\@citeb:=#2\do
	{\@citea\def\@citea{,}\@ifundefined
	{b@\@citeb}{{\bf ?}\@warning
	{Citation `\@citeb' on page \thepage \space undefined}}
	{\csname b@\@citeb\endcsname}}}{#1}}

\newif\if@cghi
\def\cite{\@cghitrue\@ifnextchar [{\@tempswatrue
	\@citex}{\@tempswafalse\@citex[]}}
\def\citelow{\@cghifalse\@ifnextchar [{\@tempswatrue
	\@citex}{\@tempswafalse\@citex[]}}
\def\@cite#1#2{{$\null^{#1}$\if@tempswa\typeout
	{IJCGA warning: optional citation argument 
	ignored: `#2'} \fi}}

\def\pmb#1{\setbox0=\hbox{#1}
	\kern-.025em\copy0\kern-\wd0
	\kern.05em\copy0\kern-\wd0
	\kern-.025em\raise.0433em\box0}

\def\fnt#1#2{\footnotetext{\kern-.3em
	{$^{\mbox{\scriptsize #1}}$}{#2}}}

\def\ps@myheadings{%
    \let\@oddfoot\@empty\let\@evenfoot\@empty
    \def\@evenhead{\slshape\leftmark\hfil}
    \def\@oddhead{\hfil{\slshape\rightmark}}
    \let\@mkboth\@gobbletwo
    \let\sectionmark\@gobble
    \let\subsectionmark\@gobble
    }
\font\tenrm=cmr10
\font\tenit=cmti10 
\font\tenbf=cmbx10
\font\bfit=cmbxti10 at 10pt
\font\ninerm=cmr9

\font\eightrm=cmr8

\def\qed{\hbox{${\vcenter{\vbox{		    
   \hrule height 0.4pt\hbox{\vrule width 0.4pt height 6pt
   \kern5pt\vrule width 0.4pt}\hrule height 0.4pt}}}$}}

\def\bsc{{\sc a\kern-6.4pt\sc a\kern-6.4pt\sc a}}  	
\def\bflatex{\bf L\kern-.30em\raise.3ex\hbox{\bsc}\kern-.14em 
T\kern-.1667em\lower.7ex\hbox{E}\kern-.125em X} 

\begin{document}
\setlength{\textheight}{8.75truein}  
\topmargin -0.25truein

\markboth{\protect{\footnotesize\it Finding low-temperature states...
}}{\protect{\footnotesize\it Finding low-temperature states...
}}

\normalsize\textlineskip

\setcounter{page}{1}



\vspace*{0.88truein}
\centerline{\bf FINDING LOW-TEMPERATURE STATES WITH PARALLEL
  }
\vspace*{0.035truein}
\centerline{\bf TEMPERING, SIMULATED ANNEALING AND SIMPLE MONTE CARLO}
\vspace*{0.37truein}

\centerline{\footnotesize J.~J.~Moreno}
\baselineskip=12pt
\centerline{\footnotesize\it Department of Physics, University
of California Davis}
\baselineskip=10pt
\centerline{\footnotesize\it Davis, CA 95616, USA}

\vspace*{15pt}
\centerline{\footnotesize Helmut G.~Katzgraber}
\baselineskip=12pt
\centerline{\footnotesize\it Department of Physics, University
of California Davis}
\baselineskip=10pt
\centerline{\footnotesize\it Davis, CA 95616, USA}

\vspace*{15pt} 
\centerline{\footnotesize Alexander K.~Hartmann}
\baselineskip=12pt
\centerline{\footnotesize\it Institut f\"{u}r Theoretische Physik,
Universit\"{a}t G\"{o}ttingen}
\baselineskip=10pt
\centerline{\footnotesize\it Bunsenstr.~9, D-37073 G\"{o}ttingen, Germany}

\vspace*{0.225truein}
\publisher{(\today)}{(\today)}

\vspace*{0.25truein}
\abstracts{
Monte Carlo simulation techniques, like simulated annealing and
parallel tempering, are often used to evaluate low-temperature
properties and find ground states of disordered systems. Here we
compare these methods using direct calculations of ground
states for three-dimensional Ising diluted antiferromagnets in a field (DAFF)
and three-dimensional Ising spin glasses (ISG).
For the DAFF, we find that, with respect to obtaining ground states, 
parallel tempering is superior to simple Monte-Carlo and to simulated
annealing. However, equilibration becomes more difficult
with increasing magnitude of the externally applied field. 
For the ISG with
bimodal couplings, which exhibits a high degeneracy, we conclude that
finding true ground states is easy for small systems, as is already
known. But finding each of the degenerate ground states with the same
probability (or frequency), as required by Boltzmann statistics,
 is considerably harder and becomes almost
impossible for larger systems.
}{}{}

\vspace*{5pt}
\keywords{Spin glasses; Random Field Systems; 
Frustrated Systems; Monte Carlo; Simulated Annealing; Parallel
  Tempering; Ground states}

\vspace*{1pt}\textlineskip
\section{Introduction}
\noindent
One of the major current challenges in statistical physics is the
study of random systems. Systems with quenched disorder
like spin glasses and random field systems\cite{young:97} have
attracted much attention during the past two decades. One is especially
interested in numerically investigating the glassy low-temperature
and ground-state behavior. In addition to the direct application,
the calculation of ground states can be used to generate low-lying exited
 states\cite{krzakala:00,palassini:00} or
domain walls in systems to test whether the given system exhibits an
ordered phase at non-zero temperature\cite{bray:84,mcmillan:84b,mcmillan1984}.
Recently, due to increased interactions
between physicists and computer-scientists, many new ground-state
techniques have
been developed\cite{opt-phys2001} in this field.

Which algorithm is suitable depends on the nature of the
problem. In computer science there are two major types of
optimization problems\cite{COM-garey79}: 
{\em easy} and {\em hard} ones. For easy
problems there exist algorithms which can find an exact optimum
while the running time of the algorithm increases only polynomially 
as a function of 
system size. For hard problems where no such algorithms are known so far,
the running times of exact algorithms grow at least exponentially with
the system size.
In statistical physics, the ground state computation of
three-dimensional
random field Ising systems is easy, while obtaining ground state for
three-dimensional Ising spin glasses is hard. Both models are
treated in this paper. 

For easy problems one can apply exact algorithms to study large
systems. But for hard problems no fast algorithms are available. Thus,
very often heuristics and approximation methods are applied. Many
widely used techniques are based on 
{\em Monte Carlo
  simulations}\cite{ANNEAL-metropolis,stauffer1993,landau2000} (MC),
which allow one to study a thermodynamical model at a given
temperature $T$. To obtain low-temperature and ground states, many
variants have been developed, such as {\em simulated
  annealing}\cite{ANNEAL-kirkpatrick} (SimA)  which
mimics slowly cooling the system and {\em parallel
  tempering}\cite{PT,Hukushima} (PT),  where a system is simulated at
several temperatures in parallel.  

In this paper we focus on parallel tempering, a method that has been recently
applied widely to spin glasses\cite{pt-applications1,ballesteros2000,katzgraber:01},
random field systems\cite{sinova2001}, clusters/molecules\cite{pt-applications2}
and various other systems\cite{pt-applications3}.
To our knowledge, so
far only comparisons of different MC techniques with PT
have been performed. This has the disadvantage that for all of these
techniques, including PT, the proof of equilibration depends on the
model and is difficult to obtain. Here we
assess how powerful PT is in generating ground states. 
We apply PT to the 
diluted (Ising) antiferromagnet in a field (DAFF) 
and to the Ising spin glass (ISG). Two questions are asked:
\begin{itemize}
\item How often are ground states obtained ?
\item Is the ground-state sampling statistically correct in degenerate
  systems? 
\end{itemize}
For the DAFF, there exists a fast exact ground-state algorithm, hence we can
study whether PT is able to obtain exact ground states and
compare to other methods. This kind of comparison has the advantage
that the correct result is known exactly and there are no problems with
equilibration for the exact ground-state calculations. 
This is different to previous
work, where only different approximation techniques have been compared.
We demonstrate here that PT is able to generate exact ground
states for the DAFF, and that is it superior to simple MC
and to SimA. Still, the method has its
limitations: with increasing size of the external field, obtaining
ground states for the DAFF becomes increasingly difficult.

We report another restriction of the method in the second part, 
where PT is applied
to the three-dimensional ISG with bimodal distribution of the
interactions. This system has a large ground-state degeneracy. For an
unbiased sampling of the ground states, each ground state must be present in
any result with the same probability. By studying small
systems that have only few different ground states and 
where one can do many independent runs, we show that
PT, while able to obtain true ground states for the ISG, 
has a harder time to get an unbiased sampling. Hence, for larger systems where
the degeneracy is larger and the algorithm is subsequently slower,
additional measures must be undertaken\cite{equi} to have an unbiased
result. This finding also applies 
to studying the low-temperature (i.e. non-zero) behavior in general.

The rest of the paper is organized as follows. First we describe
the models that are studied and the observables investigated 
in this paper. Next, the different
algorithms used are explained. In the two
following sections, the results and comparisons for the application of the
algorithms
to the DAFF and the ISG are presented. In the last section we conclude
with a summary.

\section{Models and Observables}
\noindent
To assess the efficiency of PT, SimA and simple Monte Carlo, we apply them to 
the DAFF and the ISG. In this section both
models and background information are presented.

An example for a physical realization of the DAFF is
the crystal ${\rm Fe_{x} Zn_{1-x}F_{2}}$. Besides growing practical
applications\cite{Miltenyi} of this system, many fundamental
questions are still unsolved. 
For instance, at the antiferromagnet-paramagnet transition 
some experiments\cite{Belanger1,Belanger2} find a logarithmic
divergence of the specific heat, corresponding to a 
specific heat exponent $\alpha=0$. However,
numerical simulations\cite{Nowak,RiegerYoung,Rieger}
give a negative value of $\alpha\approx-0.5$. Also a non-diverging
susceptibility was found\cite{rfimC} using exact ground-state calculations.
Therefore the nature of the transition remains unanswered.  

Two theoretical models have been proposed: the random field
Ising model (RFIM) and a direct model for the DAFF. 
Both models are expected to be in the same universality 
class\cite{Fishman,Cardy}, hence results from both should be equivalent. 
Since the RFIM is easier to simulate than the DAFF, it has been studied 
more despite the fact that the DAFF is closer to the physical system. 
In addition, disagreements about the universality have been 
found\cite{Alex,angles-d-auriac1997,sourlas1999} at $T=0$ where the critical exponents of the $\pm \Delta$ 
RFIM and the DAFF seem to be significantly different. 
Furthermore, there is still no finite-temperature study 
for large system sizes. 
This was the initial motivation to start the current work.

The DAFF is given by the following Hamiltonian:
\begin{equation}
{\mathcal H} = - J \sum_{\langle i,j\rangle} \epsilon_{i}  \epsilon_{j}
S_{i}  S_{j} - H \sum_{i} \epsilon_{i}  S_{i} \; , 
\label{hamiltonian} 
\end{equation} 
where the sum is over the nearest neighbors on a cubic lattice 
of size $L \times L \times L$. The $S_{i} = \pm 1$ represent Ising spins.
Periodic boundary conditions are applied.
As we want antiferromagnetically coupled spins, we choose
$J=-1$ and $H$ represent the externally applied magnetic field.
The dilution is introduced by randomly setting $\epsilon_{i} = 0$ for
a vacant site and otherwise $\epsilon_{i} = 1$. A realization of the
disorder is given by a set $\{\epsilon_i\}$ of fixed (i.e. quenched)
values.
In this work we use a dilution of $50\%$, i.e., we have the same number of
randomly chosen empty and filled sites, to ensure that both
occupied and vacant sites are percolating.
For the final results, an average over the disorder has to
be performed.

For the DAFF, we are studying  the physical quantities given below.
They are obtained from the recorded values of the  energy 
per spin and the staggered magnetization per spin $M$. The energy per
spin is given by the thermal and disorder average of the Hamiltonian
of the system $E(S)/N$, where 
$N$ is the number of spins ($L^{3}/2$ for 
$50\%$ dilution). 
The microscopic staggered magnetization is given by
\begin{equation}
M = {1\over N} \sum_{i=1}^{N} (-1)^{x+y+z}S_{i}\; . 
\label{magnetization} 
\end{equation}
Here, $x$, $y$, and $z$ are the spatial coordinates of the spin $S_{i}$. 
The factor $(-1)^{x+y+z}$ 
takes the order of two sub-lattices into account and ensures
that the above magnetization is unity when the system is ordered.

We evaluate the average staggered magnetization defined as
\begin{equation}
[m]_{\rm av} = [|\langle  M  \rangle|]_{\rm av} \; .
\label{magaverage}
\end{equation}
Here $\langle \ldots \rangle$ represent a thermal average 
whereas $[\ldots]_{\rm av}$ represents an average over disorder.

\begin{figure}
\centerline{\psfig{file=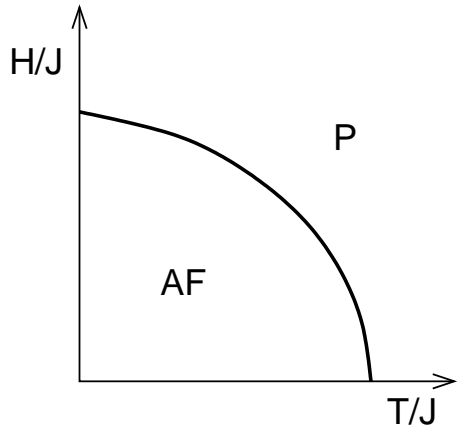,width=6cm}}
\caption{A sketch of the phase boundary of the DAFF. The
antiferromagnetic phase is denoted by 
``AF'' and the paramagnetic phase by ``P''.}
\label{figPhase}
\end{figure}

The staggered magnetization is the order parameter for the DAFF.
The canonical phase diagram 
of the DAFF in higher than two dimensions 
is shown in Fig. \ref{figPhase}.
For low temperatures and small randomness, the antiferromagnetic couplings
dominate, hence the system exhibits a long-range order. For higher
temperatures $T>T_{\rm c}(H)$, 
where the entropy dominates, or for higher fields,
where the spins are predominately aligned parallel to the field, 
the system is paramagnetic.

In addition, we evaluate the average specific heat:
\begin{equation}
 [C]_{\rm av}=N (
[\langle E^{2}\rangle]_{\rm av}-
[\langle E \rangle^{2}]_{\rm av}
)/T^{2}, 
\label{heat} 
\end{equation}
the  susceptibility
\begin{equation}
 [\chi]_{\rm av}=N (
[\langle M^{2}\rangle]_{\rm av}-
[\langle |M| \rangle^{2}]_{\rm av}
) \; ,
\label{suscep} 
\end{equation}
and the Binder cumulant\cite{binder:81}, which related to  
the ratio between the
fourth moment of the magnetization divided by the second moment squared, i.e.
\begin{equation}
 g=1/2
\left(3-
\frac{[\langle M^{4}\rangle]_{\rm av}}
{[\langle M^{2}\rangle^2]_{\rm av}}
\right) \; .
\label{binder} 
\end{equation}

Furthermore, we study the Edwards-Anderson (EA) Ising spin glass.
The Hamiltonian of the EA model is given by
\begin{equation}
{\cal H} = -\sum_{\langle i,j \rangle} J_{ij} S_i S_j \; ,
\label{ham}
\end{equation}
where the Ising spins $S_i = \pm 1$ lie on a hypercubic lattice in
three dimensions with $N=L^3$ sites, and $J_{ij}$ are nearest-neighbor
interactions chosen  according a given probability distribution.
We study the Gaussian case as it is known to have a 
unique ground state (up to
a global symmetry) as well as the bimodal case which is highly 
degenerate.\footnote{Spin glasses have been studied widely in 
statistical physics, for more
information we refer to the literature\cite{reviewSG,young:97}.
}

For the ISG, we study the ground state energy and record
histograms of how often each of the degenerate 
ground states is found by performing several simulations with the same disorder
realization but different initial spin configurations.

\section{Algorithms}
\noindent

We study the DAFF and the ISG with 5 different
algorithms. Three of them are simulation methods at finite
temperature: simple Monte Carlo, simulated annealing and parallel
tempering Monte Carlo. For the ground state calculations, we use exact
graph-theory based techniques for the DAFF and a heuristic method,
the genetic cluster-exact approximation algorithm, for the ISG.

A review on recent simulation techniques at finite temperatures
can be found in Ref. \cite{iba2001}. Here we just state the fundamental
notions behind the applied techniques.
The basic idea of the {\em Monte Carlo} approach is that a random walk in
configuration space is performed, leading to a stationary
distribution of the configurations $S$ which is the {\em Boltzmann
distribution} at fixed temperature $T$. At lower temperatures, the
algorithm freezes, hence calculating ground states is difficult,
especially for random systems like spin glasses and diluted antiferromagnets.

A more sophisticated approach to obtain low-temperature states is the
{\em  simulated annealing} (SimA) technique\cite{ANNEAL-kirkpatrick}. The
basic idea is to mimic the experimental 
techniques that are used to determine the low-temperature behavior of
a probe: the system is equilibrated at high temperature, well above
the phase where the glassy behavior occurs. Then the system is slowly
cooled down to the desired temperature. But, it remains
difficult to obtain equilibration. The reason is that one might still be
caught in local minima. Since the temperature is monotonically
decreased, it is hard to escape a minimum once the temperature is
already low.

For this reason another method has been invented, the {\em parallel
  tempering} (PT) approach\cite{PT,Hukushima}. 
The basic idea is that the temperature is not
only reduced, but it
is also allowed to raise from time to time to escape local minima of
the free energy landscape. This algorithm is realized by keeping $N_T$
different and independent configurations at temperatures
$T_{\min}=T_1<T_2< \ldots, <T_{N_T}=T_{\max}$ 
For each configuration,
several MCS are performed. Then the configurations are
allowed to exchange their positions in the list, i.e. to exchange their
temperatures. The swapping probability is chosen, such that for joint
distribution over all temperatures the joint
 Boltzmann distribution is obtained.

To assess the ability of PT and SimA to generate ground states, we
compare with exact ground states.
The ground state calculation of the DAFF can be solved in a time
that is polynomial in the number of spins.
To calculate the exact ground states at given randomness
$\{\epsilon_i\}$  and field
$H$, algorithms\cite{angles-d-auriac1997b,rieger1998,alava2001,opt-phys2001}
from graph theory\cite{swamy,claibo,knoedel} are applied. 
The calculation works by transforming the
system into a network\cite{picard1}, calculating the maximum flow
in polynomial
time\cite{traeff,tarjan,goldberg1988,cherkassky1997,goldberg1998}, and
finally obtaining the spin configuration $\{S_i\}$ from the values of
the maximum flow in the network.  The running time of the latest
maximum-flow method has a peak near the phase transition and
diverges\cite{middleton2002b,middleton2002}
 there like $O(L^{d+1})$, hence very large systems like $N=100^3$ 
can be studied easily.
Here, where we are interested in comparison with results from Monte
Carlo techniques, we restrict ourself to small systems up to $N=36^3$.

Finally, for two-dimensional spin glasses without external field with
periodic boundary conditions at most in one direction, efficient
polynomial algorithms for the calculation of exact ground states 
are available. Recently, results for systems of size $1800 \times 1800$
have been obtained\cite{palmer99}. On the other hand,
the calculation of ground states for three-dimensional ISGs  
belongs to the  class of the NP-hard problems\cite{barahona82}, 
i.e., only algorithms with exponentially increasing running time
are available. Thus, only small systems can be treated\cite{liers2000}.
The basic method used here 
is a combination of a genetic algorithm and the cluster-exact
approximation (CEA) technique\cite{alex2} which allows to obatin
ground states up to size $N=14^3$. A pedagogical
presentation of this combination of methods can be found 
elsewhere \cite{opt-phys2001}.

\section{Results for the DAFF}
\noindent
We have studied the DAFF with system size $L=12$. 
Our initial motivation has been to use PT to go to
much larger systems sizes. Simulations for sizes $L=24$ and 
$L=36$ have been performed, but only for $H=0.4$. Our experience is
that equilibration is too difficult, i.e., it is impossible to obtain
reliable results for the larger sizes. Since we are only interested in
comparing different algorithms, small sizes are sufficient.

For parallel tempering to work well one has to choose carefully a set of
temperatures to ensure that the acceptance ratios for the global moves 
are constant (as a function of $T$), preferentially near 0.5.
The number of temperatures depends on the size of the lattice and 
the lowest temperature studied.
We have used the method introduced by Hukushima\cite{Hukushima} 
where one recursively adapts the temperature set until the acceptance 
ratios are about 0.5 for all temperature pairs.

The resulting number $N_T$ of temperatures,
 the minimum and maximum temperatures,
the number of samples (i.e.\ disorder realizations) $N_{\rm samp}$, 
the total number of sweeps  $N_{\rm sweep}$ are shown in Table \ref{table1}.

\begin{figure}[ht]
\centerline{\psfig{file=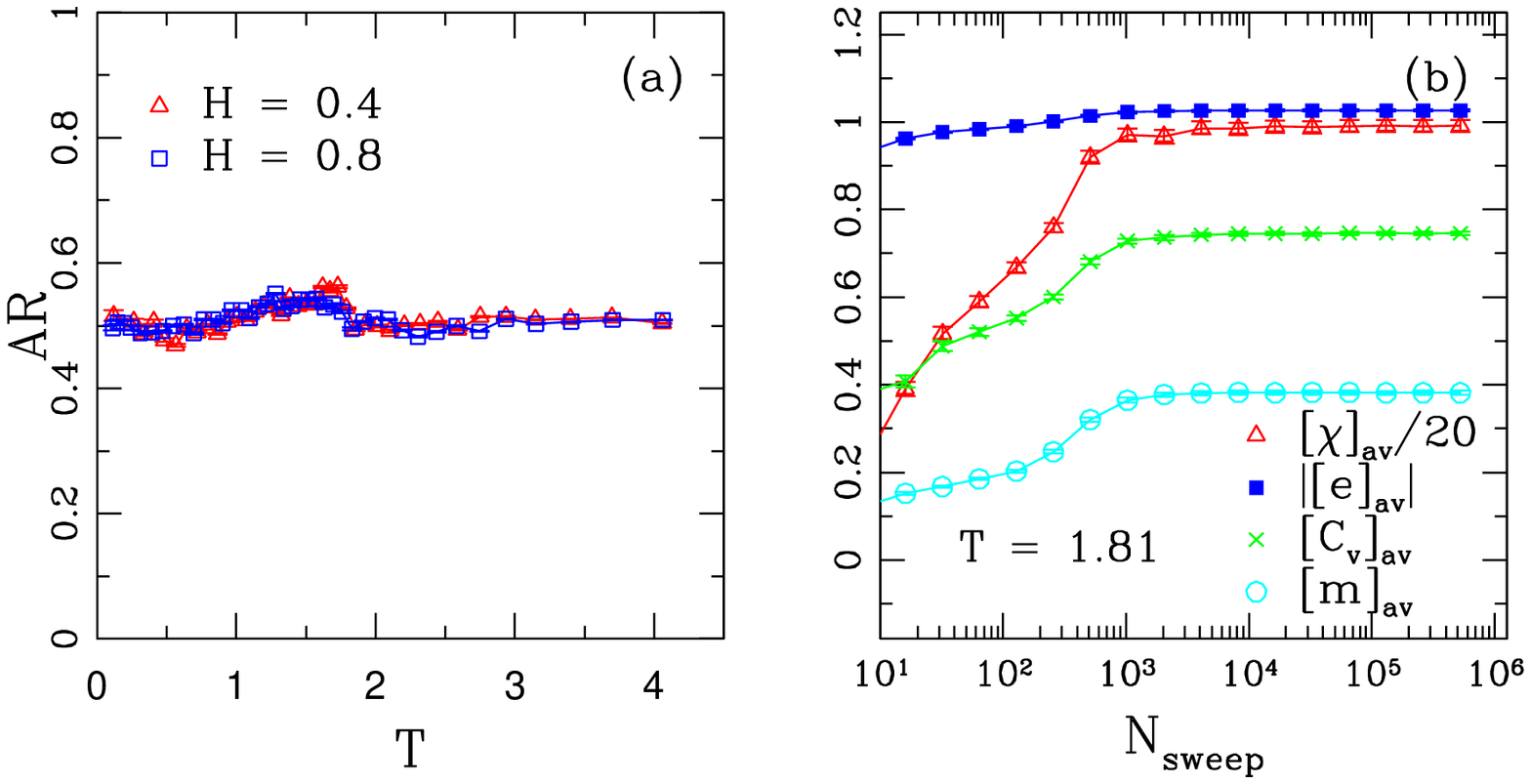,width=12cm}}
\fcaption{
Panel (a) shows the acceptance ratios
that are obtained after the simulation for the DAFF. Note that they are around $\sim 0.5$
and almost temperature independent.

In panel (b) we show different observables as a function of MCS for $L=12$
(DAFF).
One can clearly see that the staggered magnetization $[m]_{\rm av}$, 
specific heat $[C]_{\rm av}$,
absolute value of the energy per spin $|[e]_{\rm ac}|$ 
and susceptibility $[\chi]_{\rm av}$
become independent of the number of MCS at the {\it same} equilibration time.
The data is 
for $H=0.4$ and $T=1.81$. This temperature, close to the critical, shows 
the worst equilibration. The data for the susceptibility have been
divided by $20$ for better viewing.
}
\label{fig:equil}
\end{figure}

The simulations are started with a random spin configuration. 
We have checked that the acceptance ratios are almost flat
[see Fig.~\ref{fig:equil}(a)]. In Fig.~\ref{fig:equil}(b) we show
data for different observables for the DAFF as a function of Monte
Carlo steps in order to see that the system properly equilibrates
when using PT. The equilibration time $N_{\rm eq}$ is the number of
MCS needed at the lowest temperature 
such that all observables become time-independent, when
averaged over the last half of a MC run.
The numbers for $N_{\rm eq}$ given in Table \ref{table1} are only
an upper bound with an accuracy of a factor two.

\begin{table}[htbp]
\centerline{\footnotesize Parameters of the Simulation}
\centerline{\footnotesize\smalllineskip
\begin{tabular}{l  c c c c c c}\\
\hline $H$ & $N_T$ & $[T_{1},T_{N_T}]$ & $N_{\rm samp}$ & 
$ N_{\rm sweep}$ & $N_{\rm eq}$ \\
\hline 
\hline 0.4 & 40  & [0.0128, 4.2613] & 260 & $ 524288$ & $65536$  \\ 
\hline 0.8 & 42  & [0.9570, 4.2613] & 150 & $ 524288$ & $65536$  \\ 
\hline 1.2 & 47  & [0.0328, 3.7144] & 120 & $ 524288$ & $65536$  \\ 
\hline 1.6 & 50  & [0.0497, 4.0419] & 140 & $ 131072$ & $32768$  \\ 
\hline 2.0 & 46  & [0.0003, 4.0419] & 400 & $ 65536$  & $16384$  \\ 
\hline\\
\label{table1}
\end{tabular}}
\tcaption{Parameters of the simulation for the DAFF: $H$ is the externally applied
magnetic field, and $N_T$ the number of temperatures used in the parallel
tempering method. $N_{\rm samp}$ is the total number of disorder realizations,
$N_{\rm sweep}$ denotes the number of MCS per sample and temperature,
while $N_{\rm eq}$ is the equilibration time after which the
measurements have started.}
\end{table}

All these parameters are the same for the three methods used in order to 
compare the efficiency of the algorithms. This means for simple MC 
we have simulated
the system for all temperatures $T_i$ independently. 
For SimA, we have started to simulate the system with temperature
$T_{N_T}$ and cooled stepwise down to
$T_1$. Per temperature $N_{\rm sweeps}$ MCS were performed. 
This means we have {\em not} used the usual exponential cooling
schedule. Please note that the Hukushima scheme results in more
temperatures in the low temperature regime, hence the difference to the
exponential scheme is not too large.

\begin{figure}[htbp]
\centerline{\psfig{file=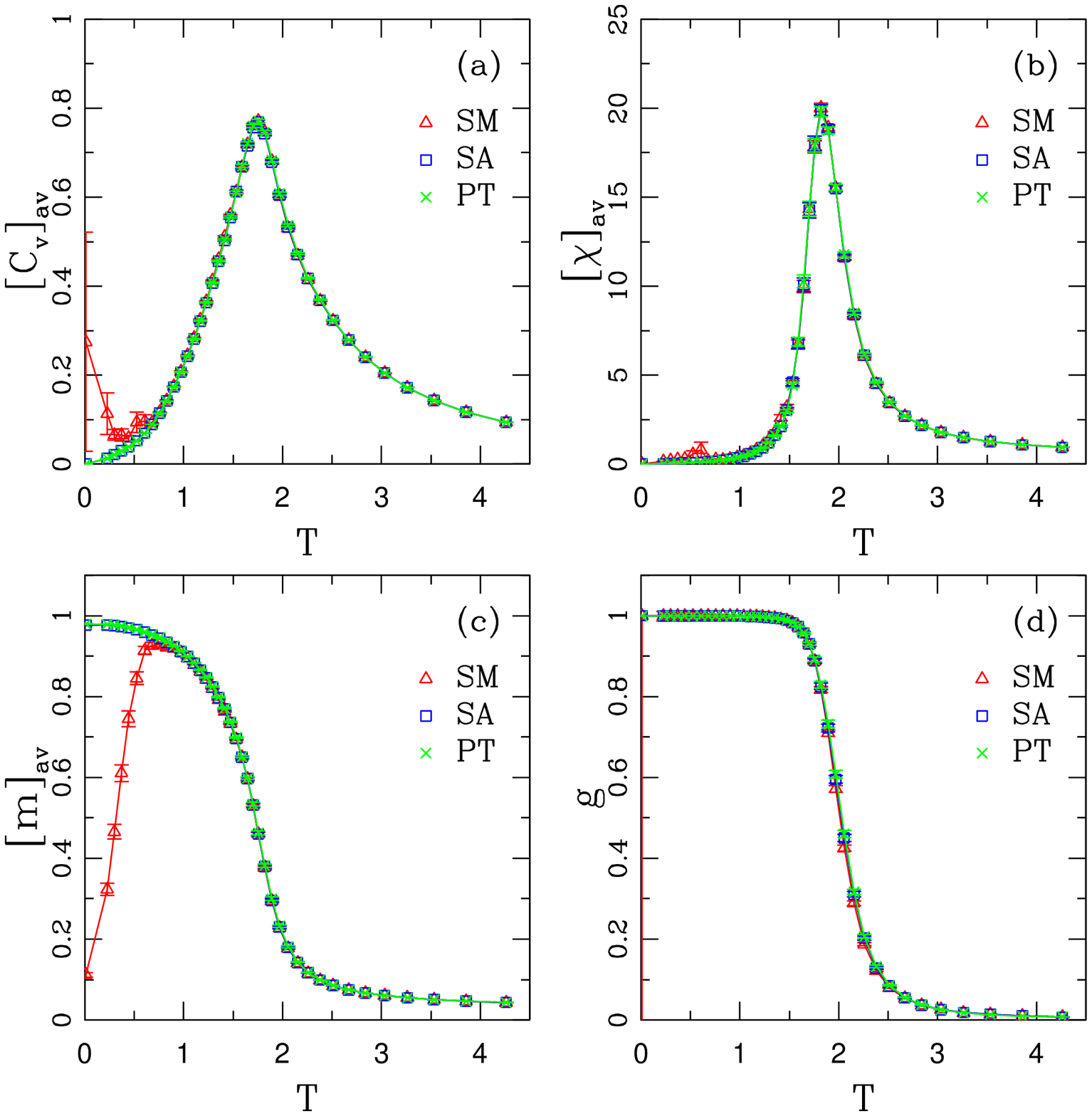,width=12cm}}
\fcaption{
Data for the different observables for $H=0.4$ using simple Metropolis (SM),
simulated annealing (SimA) and parallel tempering (PT) Monte Carlo
for the DAFF. One can see
that simple Metropolis always gives wrong results for low temperatures,
in contrast to the results from parallel tempering and simulated
annealing.
}
\label{fig:fig2}
\end{figure}

In Fig.~\ref{fig:fig2} the results at $H=0.4$ for the specific heat, the
susceptibility, the staggered magnetization and for the Binder parameter are
shown as a function of the temperature $T$ for all three methods used.
The average is taken for MC times larger than $N_{\rm eq}$ and over
all realizations of the disorder. The error bars are from sample
to sample fluctuations. One can clearly observe that for large
temperatures all three methods give the same result, while for low
temperatures the simple Metropolis approach exhibits large 
unphysical deviations, especially for the magnetization. Hence, for
the rest of the chapter we focus on the comparison between PT and SimA.

\begin{figure}[htbp]
\centerline{\psfig{file=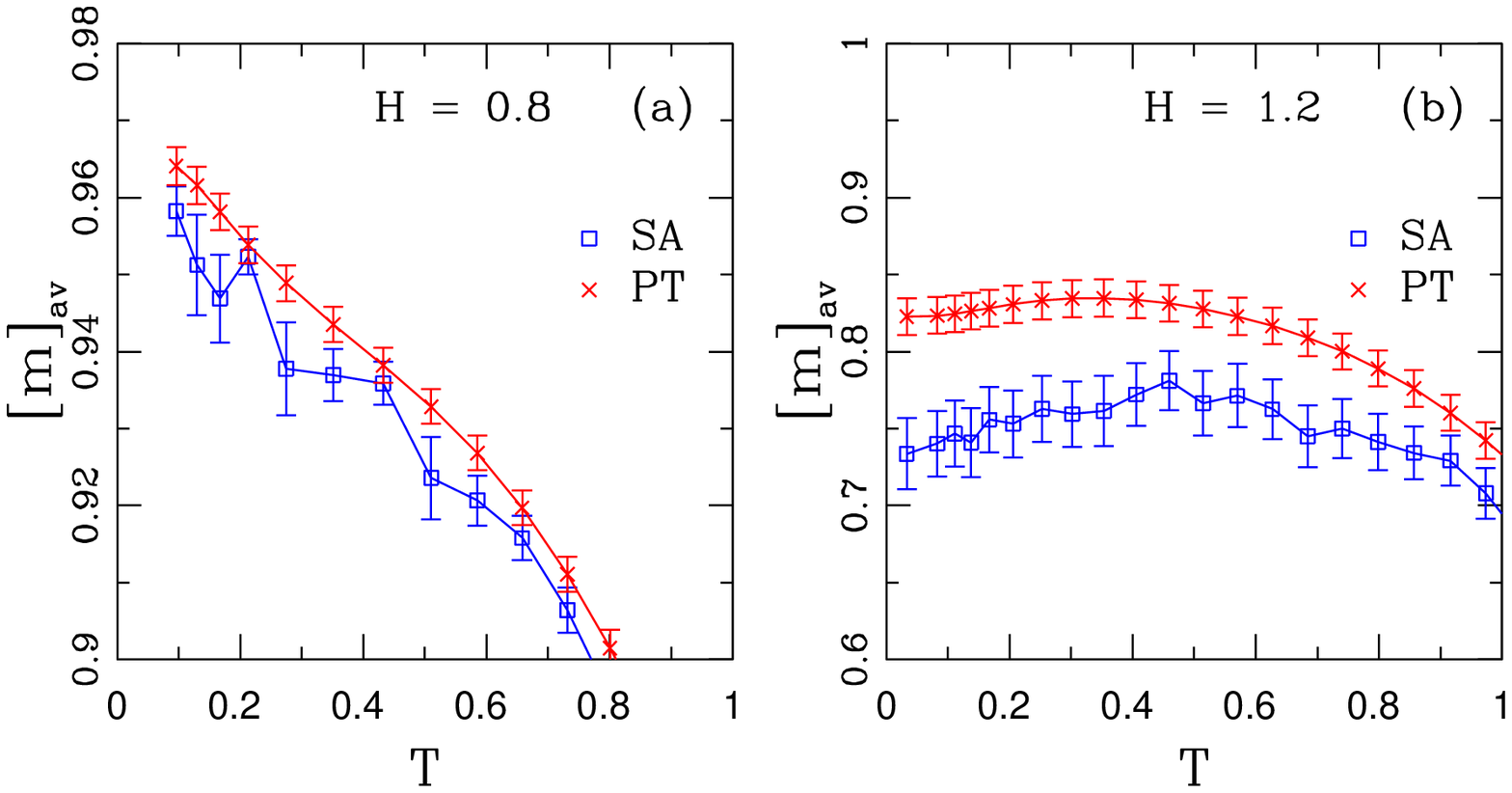,width=12cm}}
\fcaption{
Zoom on the results for the staggered
magnetization at low temperatures for $L=12$ with the applied fields
$H=0.8$ (a) and $H=1.2$ (b) for the DAFF. Although for the rest of the range the 
result is almost the same in both methods for low temperatures, the 
results from parallel tempering are more reliable.
}
\label{fig:fig3}
\end{figure}

In Fig.~\ref{fig:fig3} the staggered magnetization obtained from these two
methods is compared for larger values of the field $H=0.8$ and  
$H=1.2$, respectively.
With growing strength of the field, the deviations between
both methods increase.

\begin{figure}[htbp]
\centerline{\psfig{file=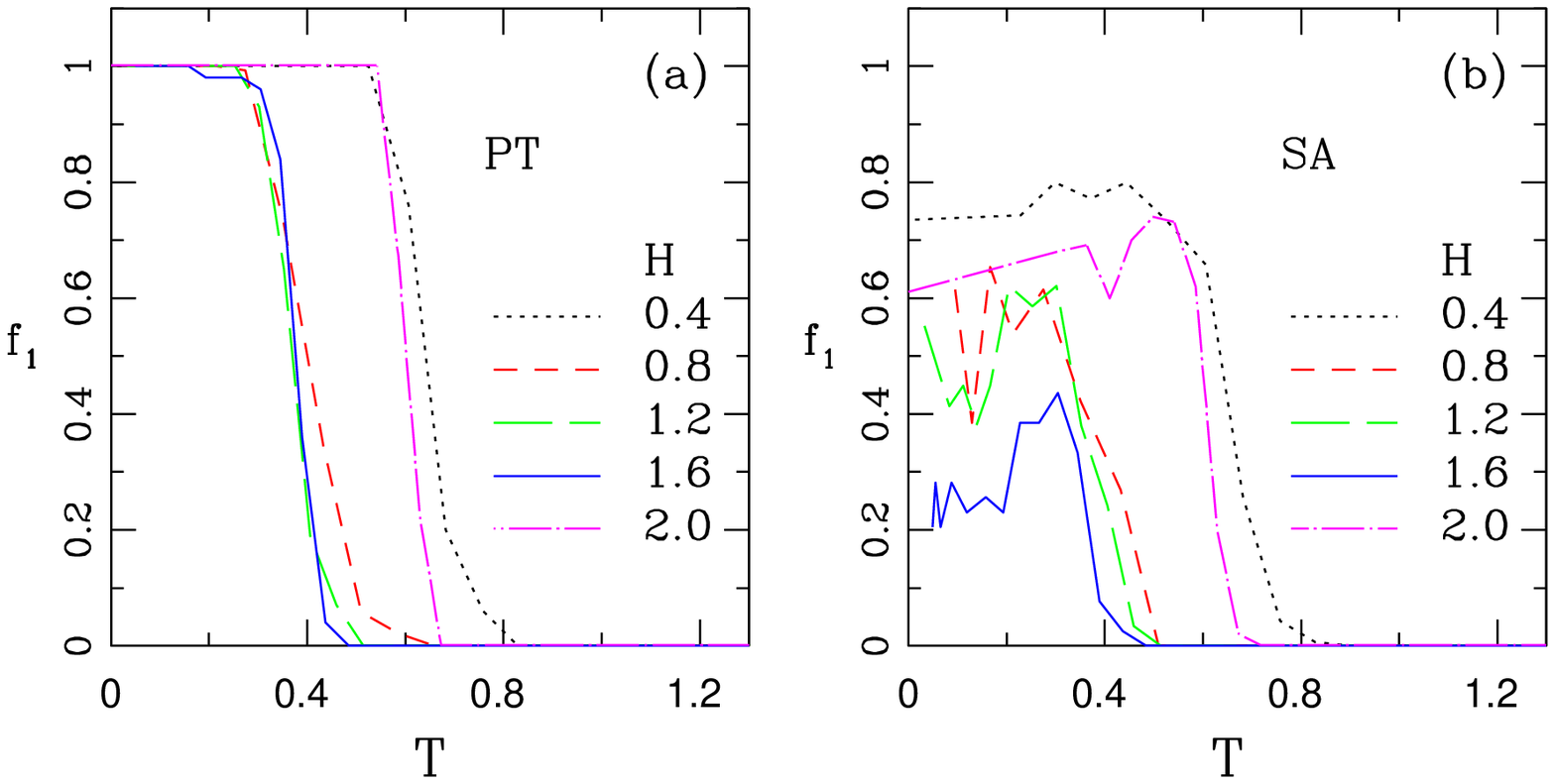,width=12cm}}
\vspace*{-1.2cm}
\centerline{\psfig{file=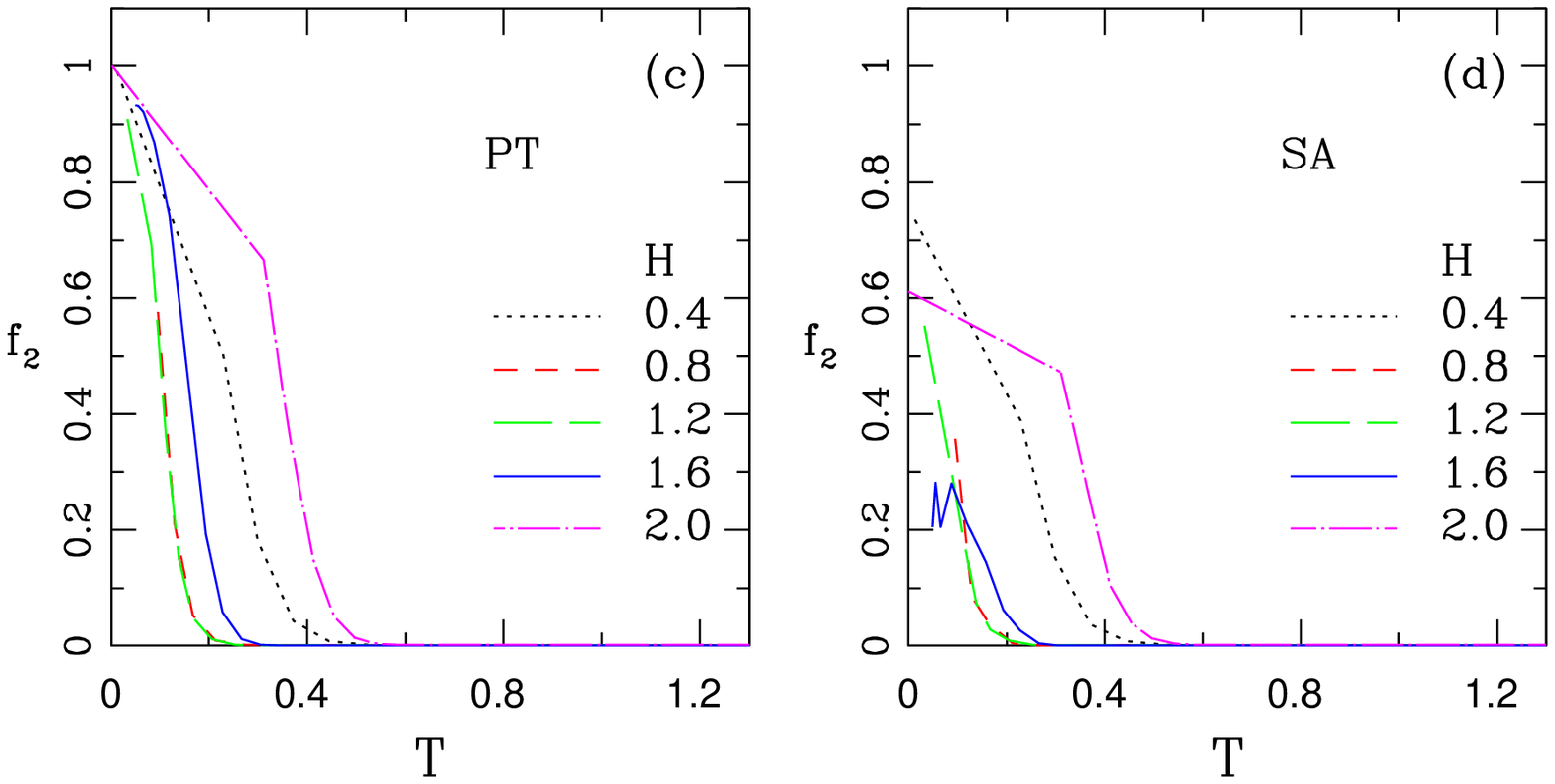,width=12cm}}
\fcaption{
This figure is a comparison between parallel tempering (PT) and
simulated annealing (SimA) for various equilibration criteria for the DAFF: the
average over of the disorder of the times the simulation finds
the ground state at least once in the range of measurements ($f_1$),
and the average over the disorder and over the range of measurements of
the times the system is in a ground state ($f_2$).}
\label{fig:f}
\end{figure}

We want to understand the cause of the difference. For this reason, we
apply the exact graph-theoretical algorithm to obtain exact ground
states and compare to the results obtained by PT and SimA. In
Fig.~\ref{fig:f} the averages $f_1$ and $f_2$ are presented. $f_1$ is
the sample average of an indicator function, which is $1$ if for a
sample at least once a ground state was found at a given
temperature, and 0 if a ground state has never been visited. 
$f_2$ is the average fraction of the simulation time a
system has been in a ground state at temperature $T$. Note that
both quantities must converge to 1 when $T\to 0$, and that 
$f_1(T)>f_2(T)$ always holds.

It is clearly visible that SimA fails in finding the true ground state
for a large fraction of all realizations. Furthermore, both $f_1$ and $f_2$
seem not to converge to 1 for $T\to 0$. With increasing size of the
external field the ability of SimA to find ground states decreases,
except for $H=2.0$ where the system exhibits a large ground-state 
degeneracy, that seems to facilitate the ground-state search.
Also, the results of SimA for $f_1$
exhibits strong fluctuations, another indication how unreliable the
result is.

On the contrary, the data for PT behaves as expected, both $f_1$ and
$f_2$ converge to 1 for $T\to 0$. This in turn shows that it is very
likely that indeed PT has obtained the true thermodynamic average, hence it is
superior to SimA (and simple MC). Again, finding ground states
becomes more difficult with growing size of the field, except the highly
degenerate case $H=2.0$. A possible explanation for this fact can be
the existence of temperature ``chaos''. This means that for
neighboring temperatures, the free-energy landscape looks very
different. Because PT relies on the fact that neighboring temperatures
yield a similar behavior, the existence of chaos would prevent a quick
equilibration. So far mostly spin glasses have been investigated
regarding the existence of chaos.
For zero field no definite answer on the existence of
chaos has been given, but
chaos seems to be weak and/or difficult to observe
\cite{billoire2000,billoire2002,takayama2002,aspelmeier2002}. 
For the RFIM, which is very similar to the DAFF, chaos has been
studied only for small random perturbations of the bonds. In this case,
for the three-dimensional model, chaos seems to be marginal\cite{alava1998}. 
For the case of temperature chaos in the DAFF, 
we believe that due to the additional field term in the Hamiltonian,
the effect
might be much larger for systems in an external field. Our result that finding
ground states becomes increasingly difficult with $H>0$ indicates that this might
be indeed the case.

Note that using this comparison of the ability to obtain ground
states, the deviations between the different algorithms are clearer
than by simply comparing physical quantities. Especially
for low values of the applied field $H$, PT and SimA give comparable results
(cf.~Fig.~\ref{fig:fig2}). The reason is that for low values of $H$,
the physical quantities for ground states and excited states do not
differ to much, hence the failure of SimA is not revealed.

In this section we have demonstrated that parallel tempering
 is superior to simple MC simulations and simulated annealing,
for moderate system sizes.
The reason being that PT is much better in sampling the
low-temperature region, especially the ground states, as demonstrated
by comparison with exact ground states obtained by a graph theoretical
algorithm. In the next section we compare PT and SimA for another
model where no fast exact ground-state algorithms are available.

\section{Ground-state Statistics for the 3D Edwards-Anderson Ising Spin Glass}
\noindent
In this brief section, we apply PT and SimA to the
three-dimensional Edwards-Anderson (EA) Ising spin glass with a Gaussian
respectively bimodal distributions of the interactions. We 
show that PT is able to calculate true ground states
and that we obtain the statistics of the ground
states for the degenerate bimodal systems. For the results given
below, we have also
 checked with the genetic cluster-exact
approximation method explained above that indeed ground states are found.

\begin{figure}[htbp]
\centerline{\psfig{file=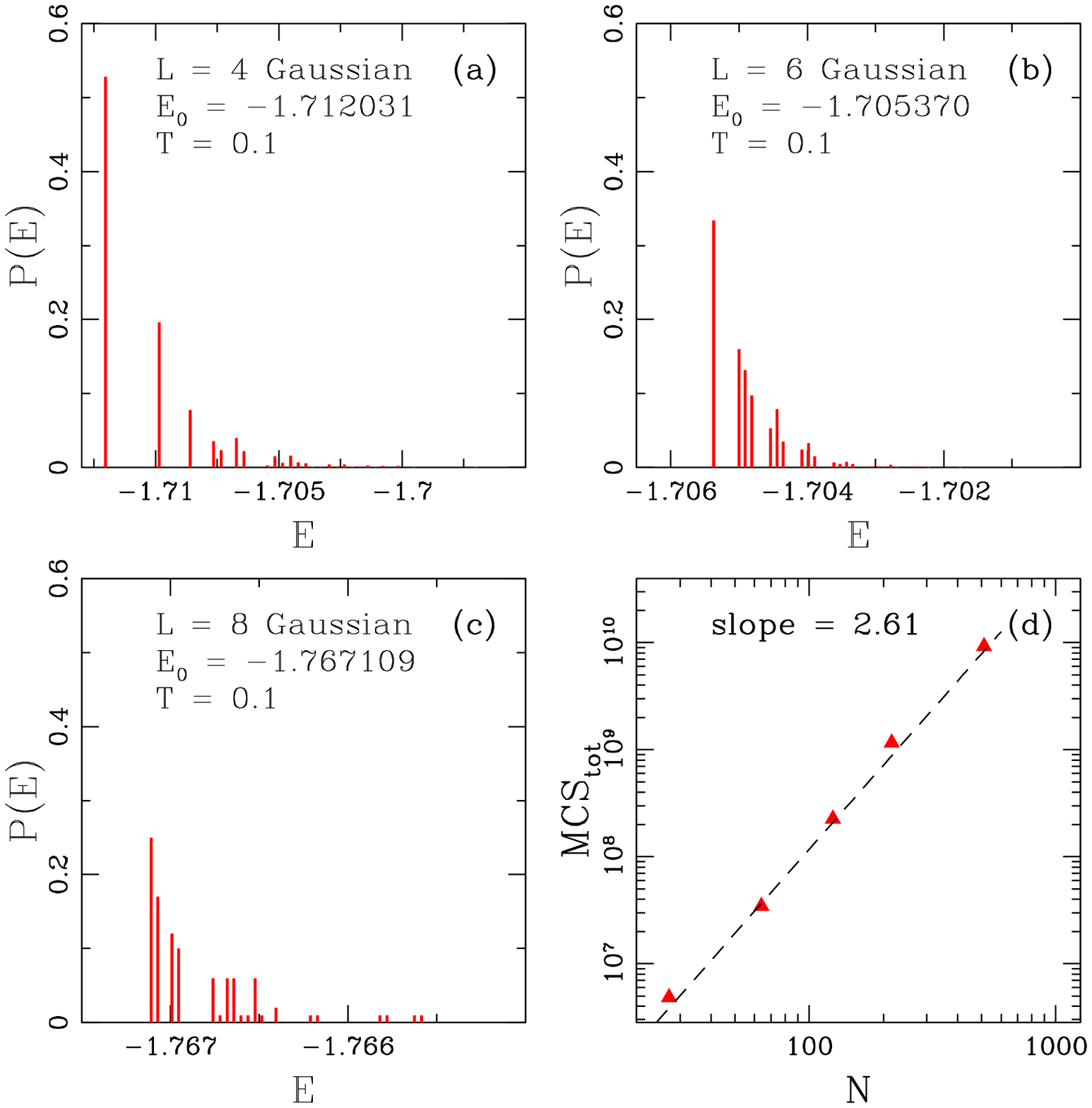,width=12cm}}
\fcaption{ 
Panels (a) -- (c) show data for the probability to
find a given energy $P(E)$ for the EA model with Gaussian bonds
for three single realizations 
of size $L = 4$, $6$, and $8$. One can
see that the ground state energy $E_0$ is the most
probable state for $T = 0.1$. Panel
(d) shows the total number of MCS needed to 
(including the number of temperatures used and spins in the system)
evaluate one ground-state realization as a function of the number
of spins $N$. A clear power-law increase with a power of $2.61$ can be
seen.
}
\label{fig:ea}
\end{figure}

To determine the energy of the system for a given set of bonds, we
equilibrate for 
a given number of MCS, in this case $10^5$ per
temperature set, and take a snapshot of 
the system configuration.
Figures \ref{fig:ea} (a) -- (c) show data for the probability to
find a given energy $P(E)$ at temperature  $T \approx 0.1T_c$
($T_c \approx 0.95$)\cite{3dtc} of three single realizations
with Gaussian distributed bonds
with $L = 4$, $6$, and $8$. We choose the Gaussian case
first as we have a strong test\cite{katzgraber:01} to see
when the system has equilibrated. We see that for
low enough temperatures, in this particular case,
the ground state $E_0$ is the most probable state for the system.
For the simulations we have used a set of 18 temperatures in a range
$[0.1,2]$. 

We have again measured the equilibration time\cite{katzgraber:01}
$N_{\rm eq}$
i.e. the number of MCS needed such that all measured quantities become
independent of the simulation time. Interestingly, it increases 
for this small system sizes 
with a power law as a function of the number of spins, as can be seen in
Fig.~\ref{fig:ea} (d). Since the spin-glass ground-state calculation
is computationally a hard problem\cite{barahona82}, we expect that for
larger system sizes, the running time increases exponentially with
system size. This has been observed for algorithms that are
especially designed to compute spin glass ground
states\cite{SG-alex-stiff,SG-alex-4d}. Note that PT is slower
than those special algorithms, but on the other hand 
is applicable to a variety of models due to its
general structure and very simple implementation.

\begin{figure}[htbp]
\centerline{\psfig{file=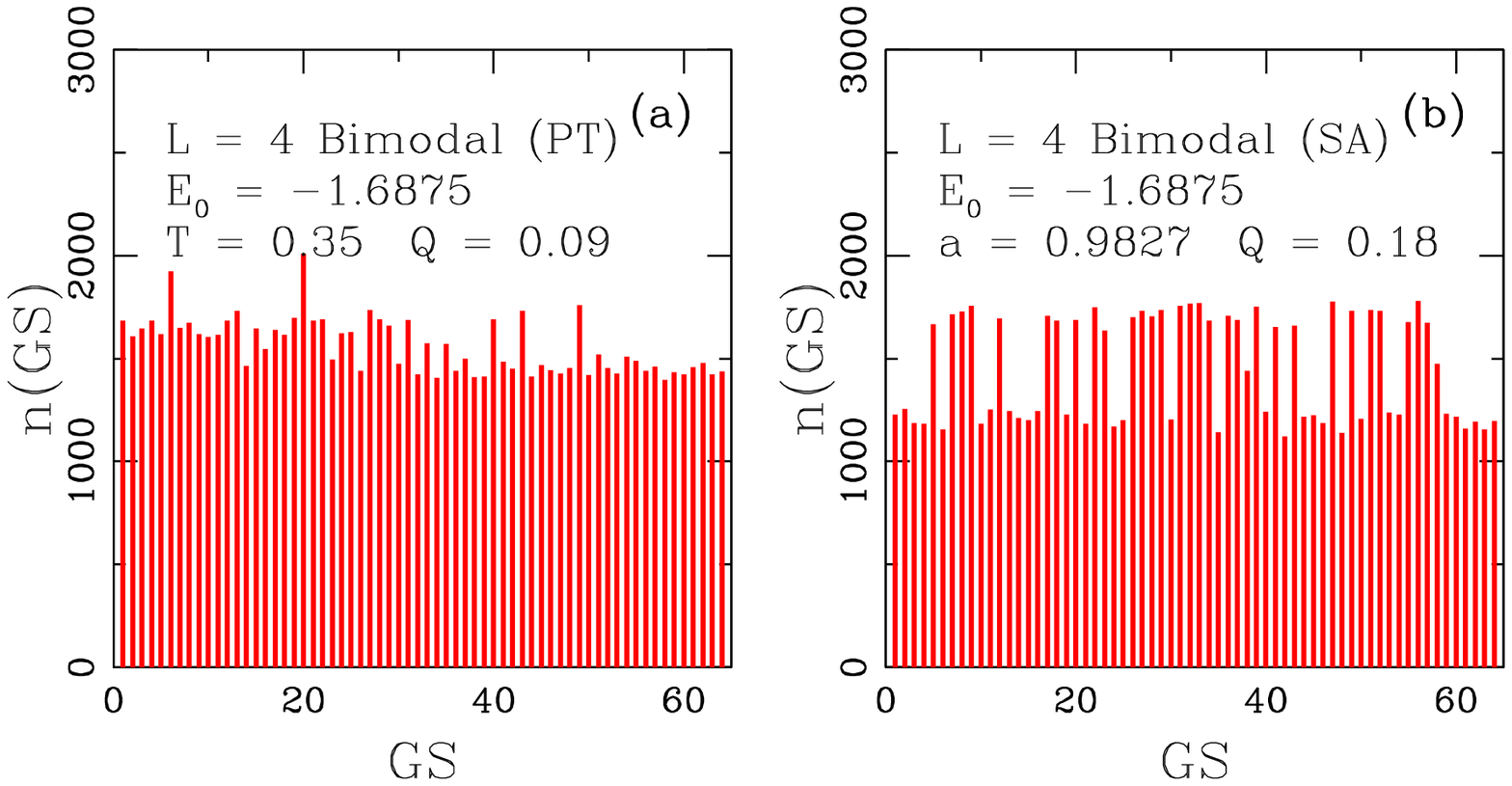,width=12cm}}
\fcaption{
Panel (a) shows the number of hits for a specific ground-state configuration
for the bimodal EA spin glass for $T = 0.35$ using parallel
tempering (PT) Monte Carlo. Data for $N_{\rm sweep} = 10^2$ MCS,
$N_T = 10$ temperatures 
and $N_{\rm runs} = 10^5$ independent runs.
For our particular choice
of bonds, we find 64 distinct ground-state configurations. Note that
the distribution of states, while not completely flat, gives almost
equiprobable ground-state configurations with $Q = 0.09$. This
is in contrast to panel (b) where we show data obtained from simulated
annealing (SimA) where $Q = 0.18$ for a cooling rate $a = 0.9827$, which
corresponds to $10^3$ MCS.
}
\label{fig:fig5}
\end{figure}

\begin{figure}[htbp]
\centerline{\psfig{file=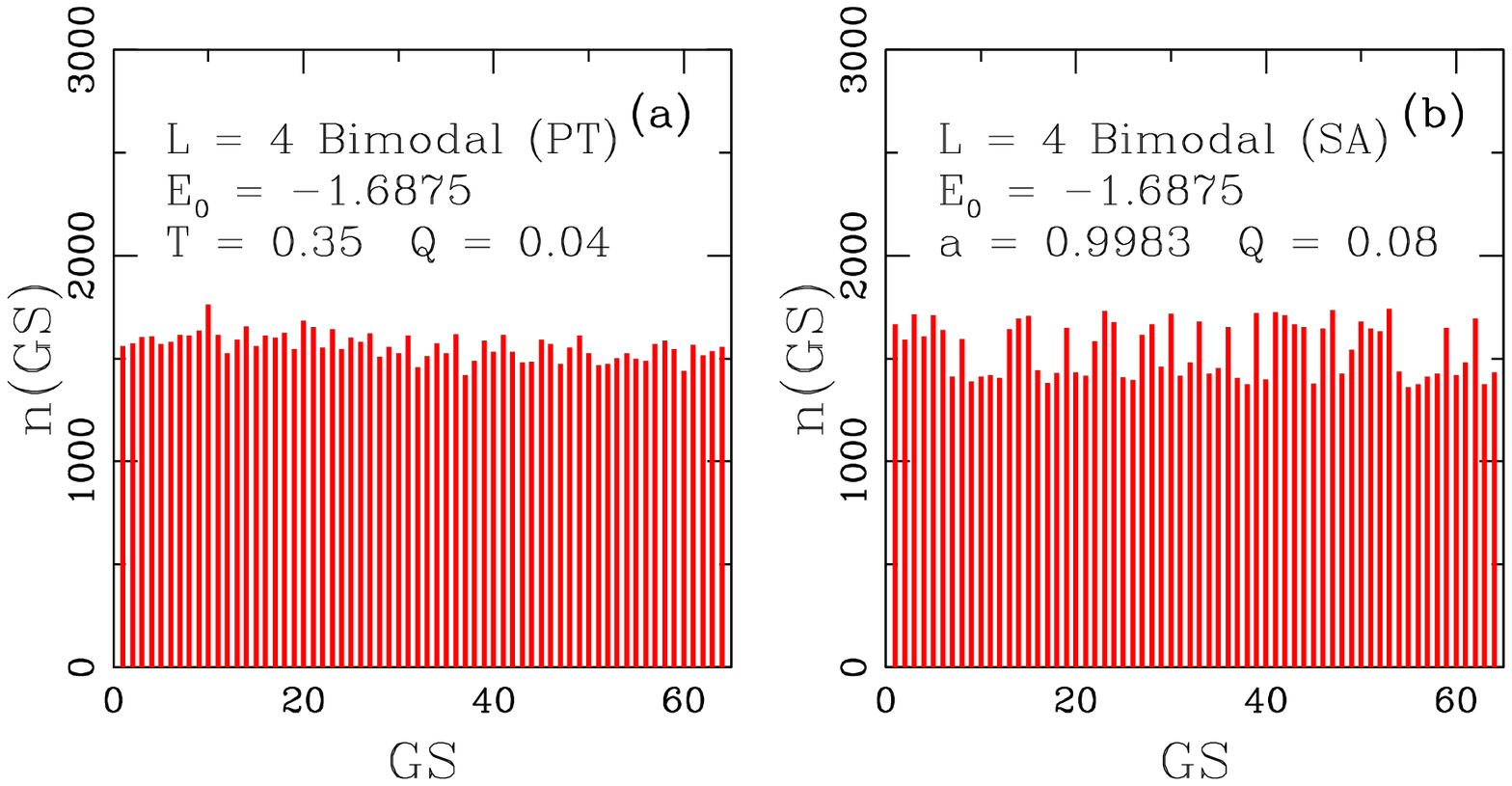,width=12cm}}
\fcaption{
Same as Fig.~\ref{fig:fig5} but for $10^4$ MCS. One can
see that for parallel tempering (PT) $Q = 0.04$ whereas for simulated
annealing (SimA) $Q = 0.08$ for a cooling rate $a = 0.9983$ which corresponds to
$10^4$ MCS.
}
\label{fig:fig6}
\end{figure}

To further illustrate the use of PT and SimA to determine ground states
for spin-glass systems, we have applied them to the highly degenerate
bimodal ISG. For the Gaussian case above, we have equilibrated the
system at low temperatures, then we have checked that indeed ground
states are found. Now we proceed in a different way. We make the
simulations for different number of MCS and investigate directly
whether ground states are obtained by comparing with the results from
the genetic CEA approach. If indeed ground states are found, 
we furthermore check if
finding ground states automatically means that the system is
equilibrated. This is tested by measuring the frequency different
degenerate ground states are found. 

In this case we have used a set of $N_T=10$
temperatures in the range $[0.35,2]$. 
We present in Fig.~\ref{fig:fig5} (a) the
ground-state statistics for a specific choice of interaction bonds for the
bimodal case calculated with PT for 
$N_{\rm sweep} = 10^2$ MCS per replica,  and 
$N_{\rm runs} = 10^5$ independent runs. The system we have studied has
64 distinct  ground states.  
For a thermodynamically correct sampling, each ground state has to be
found with the same frequency\cite{SG-sandvic1999}. 
Hence, equilibration can
be tested unambiguously by counting how often each ground state is found.
As we can see, 
the distribution of ground states is not completely flat, although
the algorithm often finds a ground state, hence is already very
efficient. At least
the results are better than those obtained by means of simulated annealing
[see Fig.~\ref{fig:fig5} (b)]
for the same total number of MCS ($10^3$) and within the
same range of temperatures, but with an exponential temperature
schedule $T_{n+1}=aT_n$

In Fig.~\ref{fig:fig6} we show the same results as in Fig.~\ref{fig:fig5}
but for $10^4$ MCS. One can clearly see that fluctuations are reduced. 

\begin{table}[htbp]
\centerline{\footnotesize Ground-state statistics for $T = 0.35$}
\centerline{\footnotesize\smalllineskip
\begin{tabular}{l c c c c c c}\\
\hline $N_{\rm sweep}$ & $N_{\rm T}$ & $N(E_0)$ & $e(GS)$ & $\sigma(GS)$ & $Q$ \\
\hline
\hline $10^1$ & 10 & $68987$  & $1061.34$ & $490.04$  & $0.46$ \\
\hline $10^2$ & 10 & $99973$  & $1538.05$ & $133.91$  & $0.09$ \\
\hline $10^3$ & 10 & $99972$  & $1538.03$ & $ 66.22$  & $0.04$ \\
\hline $10^4$ & 10 & $99970$  & $1538.00$ & $ 63.78$  & $0.04$ \\
\hline $10^5$ & 10 & $99976$  & $1538.09$ & $ 66.87$  & $0.04$ \\
\hline\\
\end{tabular}}
\tcaption{Ground-state statistics for $T = 0.35$ in the bimodal EA model.
For increasing number of MCS ($N_{\rm sweep}$), the number of
times we reach the ground state $N(E_0)$ increases. Also shown are the
mean number of times a specific ground state configuration is reached
[$e(GS)=N(E_0)/64$], as well as its standard deviation [$\sigma(GS)$]
and relative fluctuations [$Q = \sigma(GS)/e(GS)$].
Note the number of MCS quoted is 
per replica. Hence, the total
number of MCS of the simulation is given  by $N_{\rm
T}\cdot N_{\rm sweep}$, where $N_{\rm T}$ is the number of temperatures
(replicas) used. Data for $N_{\rm samp} = 10^5$ disorder realizations.}
\label{table:ea1}
\end{table}

\begin{table}[htbp]
\centerline{\footnotesize Ground-state statistics for $T = 0.65$}
\centerline{\footnotesize\smalllineskip
\begin{tabular}{l c c c c c c}\\
\hline $N_{\rm sweep}$ & $N_{\rm T}$ & $N(E_0)$ & $e(GS)$ & $\sigma(GS)$ & $Q$ \\
\hline 
\hline $10^1$ & 10 & $10322$  & $ 158.80$ & $ 70.27$  & $0.44$ \\
\hline $10^2$ & 10 & $87951$  & $1353.09$ & $103.92$  & $0.08$ \\
\hline $10^3$ & 10 & $88446$  & $1360.71$ & $ 50.47$  & $0.04$ \\
\hline $10^4$ & 10 & $88298$  & $1358.43$ & $ 59.10$  & $0.04$ \\
\hline $10^5$ & 10 & $88673$  & $1364.20$ & $ 57.48$  & $0.04$ \\
\hline\\
\end{tabular}}
\tcaption{
Same as Table \ref{table:ea1} but for $T = 0.65$.
}
\label{table:ea2}
\end{table}

In Tables \ref{table:ea1} and \ref{table:ea2} we show the dependence of the
number of times we reach the ground state as a function of MCS for
$T = 0.35$ and $0.65$, respectively. (For the bimodal case
$T_c \approx 1.15$)\cite{ballesteros2000,bimtc}. 
While the
system is known to be properly equilibrated for $N_{\rm sweep} = 10^5$ 
MCS (per temperature set) for $T = 0.35$,
we see that for shorter equilibration times we still reach the ground state 
several times. We also calculate the mean number of times 
a ground state is reached [$e(GS)$] and estimate a relative fluctuation $Q$ by 
dividing the standard deviation [$\sigma(GS)$] by the mean. 
The distribution
of ground states is fully equiprobable in the limit $Q \rightarrow
0.025$. This can be understood as follows: drawing
from $G$ numbers randomly (in this particular 
case we have $G = 64$ ground states) with all being equiprobable with
probability $P = 1/G$ follows a binomial distribution. If we repeat this
$N$ times we obtain a standard deviation $\sigma =
\sqrt{Np(1-p)}$ and expectation value $e = Np$. 
This means that $Q = \sqrt{(p - 1)/(Np)}$, and for our case with
$G = 64$ and $N = 10^5$ ($=N_{\rm runs}$) we obtain 
$Q_{\rm th} = 0.025$. 

As we can see, fluctuations decrease dramatically for both
temperatures when  the number of MCS is increased but saturates at $Q
= 0.04$. We see that we are rather
close to the theoretical value showing that while the
fluctuations are very small it would require prohibitively longer
equilibration times to find $Q_{\rm th} = 0.025$.

Here again, parallel tempering shows an improvement over simulated annealing
when trying to determine ground states and ground-state statistics. In
particular, we see that for small systems 
with moderate effort we can determine the ground state
of a spin system with equiprobable ground-state statistics. 
The effort increases
considerable for larger systems. There, not only the computational effort to
find some ground states is already significantly larger, but also the
ground-state degeneracy grows exponentially with system size. For example, a
system of size $N=6^3$ can have $10^{16}$ ground states, and it
is impossible to generate a histogram as done for small system
sizes. Hence, when evaluating ground-state properties,
one {\em cannot} take the results from MC-like
simulations directly. Finding ground states in equilibrium is more difficult than
just finding ground states. Therefore, one has to use long simulation
times at low temperatures and wait for measurable quantities to become
time-independent, or additional algorithms\cite{equi} 
must be applied to guarantee a correct sampling.

\section{Summary and discussion}
\noindent

In conclusion, we have studied the three-dimensional diluted 
Ising antiferromagnet in a magnetic field and the Edwards-Anderson
Ising spin-glass using parallel tempering Monte Carlo, simulated
annealing and simple Metropolis Monte Carlo. We have used the generation of ground
states as an assessment tool to compare these three techniques. 
We find the following:
\begin{itemize}
\item PT is superior to simulated annealing and simple Monte Carlo: More
  ground states are generated at very low temperature, where the system
  is preferentially in a true ground state for small systems.
\item For systems in an external field, generating ground states
  becomes more difficult, except if the degeneracy is very large.
  We have also performed similar tests for the ISG where we have observed the
  same behavior (not shown here). A possible explanation could be the
  existence of
  stronger temperature ``chaos'' in comparison to the zero field case.
\item Although PT is superior to the other methods studied, generating ground
  states with the correct statistics is more difficult than simply
  generating ground states at all. For larger than tiny systems, 
 it is impossible to assess directly whether the sampling is correct. 
 Either 
 additional methods must be applied or very long MC simulation times
 must be accepted.
\end{itemize}

We conclude:
for system where exact algorithms are
available that run in polynomial time, like for the DAFF,
it is recommended not to use a simple Monte Carlo
approach (like PT) for the ground state calculation. 
The reason is that one can treat much larger system sizes
by using the exact techniques than it is possible with MC, while MC 
does {\em not guarantee} true ground states.
If one is interested not only in the exact zero-temperature behavior
but in the non-zero low-temperature behavior, one should use a combination of
the exact techniques and PT. When only applying Monte-Carlo techniques
like PT, it seems unlikely that for the DAFF it is possible to
equilibrate large systems (such as $100^3$ spins). In this aspect the DAFF 
is of similar complexity as spin glasses. This seems strange,
because for the ground-state problem, the DAFF is much easier 
to simulated than ISGs. 

For the ISG and other problems, where no fast exact algorithms are
available, currently PT is the best choice for an algorithm that 
is general purpose and simple to implement. But, as we have seen,
one cannot say that PT is orders of magnitudes faster than standard
algorithms used. It suffers from the same problems as SimA or simple MC, 
but equilibrates faster. Care has to be exercised before results
are believed. As we have seen it is still easily possible that the
true equilibrium behavior is not obtained. Several independent
equilibration tests should always be applied.

\section{Acknowledgments}
\noindent
We would like to thank A.P. Young for many discussions, critically
reading the manuscript and hosting us at the University of California
Santa Cruz where this cooperation started. Furthermore, we appreciate
 K.-V. Tran spell- and grammar checking the manuscript.
JJM would like to thank to A.~Slepoy for fruitful discussions
and to Scalettar-Zimanyi Group for their hospitality. Besides, he would like 
to acknowledge financial support from Ministerio de Educaci\'{o}n, 
Cultura y Deporte de Espa\~{n}a under a F.P.I.~grant.  
HGK would like to acknowledge financial support
from the National Science Foundation under grant No.~DMR 9985978.
AKH obtained financial support from the DFG (Deutsche 
Forschungsgemeinschaft)
under grants Ha 3169/1-1 and Zi 209/6-1.
This research was supported in part by NSF cooperative agreement ACI-9619020
through computing resources provided by the National Partnership for Advanced 
Computational Infrastructure at the San Diego Supercomputer Center.
We would like to thank the University of New
Mexico for access to their Albuquerque High Performance Computing Center. 
This work utilized the UNM-Alliance Los Lobos Supercluster.

\nonumsection{References}

\end{document}